\documentclass[english,pra,twocolumn,superscriptaddress]{revtex4-1}
\usepackage{amsmath}
\usepackage{amssymb}
\usepackage{graphicx}
\usepackage{times}

\begin{document}
\title{Pulse-qubit interaction in a superconducting circuit under frictively
dissipative environment}
\author{Yibo Gao}
\affiliation{College of Applied Sciences, Beijing University of Technology, Beijing,
China}
\author{Shijie Jin}
\affiliation{College of Applied Sciences, Beijing University of Technology, Beijing,
China}
\author{Yan Zhang}
\affiliation{Institute of Applied Physics and Materials Engineering, University
of Macau, Macau, China}
\author{Hou Ian}
\email{houian@um.edu.mo}

\affiliation{Institute of Applied Physics and Materials Engineering, University
of Macau, Macau, China}
\affiliation{Zhuhai UM Science \& Technology Research Institute, Zhuhai, Guangdong,
China}
\begin{abstract}
Microwave pulses are used ubiquitously to control and measure qubits
fabricated on superconducting circuits. Due to continual environmental
coupling, the qubits undergo decoherence both when it is free and
during its interaction with the microwave pulse. As quantum logic
gates are executed through pulse-qubit interaction, we study theoretically
the decoherence-induced effects during the interaction, especially
the variations of the pulse, under a dissipative environment with
linear spectral distribution. We find that a transmissible pulse of
finite width adopts an asymmetric multi-hump shape, due to the imbalanced
pumping and emitting rates of the qubit during inversion when the
environment is present. The pulse shape reduces to a solitonic pulse
at vanishing dissipation and a pulse train at strong dissipation.
We give detailed analysis of the environmental origin from both the
perspectives of envelope and phase of the propagating pulse.
\end{abstract}
\maketitle

\section{Introduction}

Superconducting qubit circuits have become a major platform for testing
quantum computation protocols~\citep{clarke08} and quantum optical
effects~\citep{jqyou11,ian10}. As a solid-state device fabricated
on semiconductor substrates, a superconducting circuit suffers from
noises originated from multiple sources such as the underlying dielectric~\citep{martinis05}
and the bias input~\citep{martinis03}, most notably in form of $1/f$
noise~\citep{bialczak07,yoshihara06}. The noises induce decoherences,
in terms of both transverse dephasing and longitudinal relaxations,
in the qubits~\citep{ithier05}. Hence, succesful demonstrations
of optical effects and computational operation, such as state readouts,
often become contests against the decoherence times~\citep{mallet09}.
A quantitative understanding of decoherence~\citep{yu17} becomes,
therefore, essential in the further development of superconducting
circuits.

Throughout the years, multiple studies have been devoted to the descriptions
of decoherences on this solid-state system~\citep{tian02,jqyou07,mcdermott09}.
In particular, Ref.~\citep{tian02} models a SQUID loop induced decoherence
during readout, based on a noise spectrum framed on the Leggett model
of infinitesimally-spaced linear resonators\citep{caldeira83,leggett84}.
Ref.~\citep{jqyou07} estimates the relaxation of a standalone qubit
directly from Fermi golden rules. Few studies actually consider the
decoherence induced on a superconducting qubit during its interaction
with a microwave pulse. A correct modeling of the decoherence induced
during the course of interaction would benefit the designs of entanglement
operations~\citep{wei06,neeley10} and projective measurements~\citep{mallet09}
on qubits, which are carried out through coupling them with definite
types of microwave pulses.

Since a resonant pulse forms time-dependent dressed states with the
qubit, which give rise to different decay channels than those of the
qubit bare states. the qubit relaxations are highly dependent on the
variations of the envelope and phase of the traveling microwave pulse.
Inversely, the change incurred on the envelope and phase after the
pulse propagates through the qubit are highly dependent on the decoherence
of the qubit, which are often ignored in the studies related to resonant
pulses. Here, we use an adiabatic master equation to register the
environmental influence on the qubit during the course of interaction
and compute how a given dissipative environment would affect the output
pulse in both its envelope and phase after it is applied to the qubit.

In line with superconducting circuit systems, which contain the qubits
that are made up by Josephson junction barriers~\citep{martinis05},
the environment is assumed to be frictive~\citep{caldeira81}, where
the barriers are considered Ohmic interfaces. In other words, its
spectral density $J(\omega)$ is assumed linear in its frequency dependence~\citep{widom82}.
Under such an assumption, we find that resonant pulses behave differently
during their propagation through the qubit, depending on the linear
scale factor embodied in the spectrum density. Modeling the qubit
evolution through a microscopic master equation and the pulse travel
through Maxwell equations, we obtain the limiting solution for a vanishing
scale factor to be a solitonic pulse, which adopts a symmetric hump
shape. When the scale factor becomes large to represent a strong dissipative
environment, the solution approaches another asymptotic limit that
associates with a continuous pulse train.

Between these two limits, solutions to the coupled model equations
admit pulse envelope shapes with multiple humps of various heights
and monotonic increase in phase with various rates, all of which depend
on the magnitude of dissipation. Pulses are only admissible with multiple
peaks because the qubit and the environment are of competing nature
in the absorption and the re-emission of microwave pulses. We analyze
quantitatively how the enveloped pulse area is determined by this
competing nature in Sec.~\ref{sec:Propagation}. The analysis is
based on the solution to the coupled Maxwell-master equations derived
from an adiabatic pulse-qubit interation model we explain in Sec.~\ref{sec:model}.
We also analyze the variation of the pulse phase in Sec.~\ref{sec:Phase}
before presenting the conclusions in Sec.~\ref{sec:Conclusions}.

\section{Pulse-qubit interaction\label{sec:model}}

The interacting system of a qubit and an incident pulse is described
by the time-dependent Hamiltonian ($\hbar=1$)
\begin{equation}
H_{S}(t)=\frac{\omega_{z}}{2}\sigma_{z}-\mu E(t)(\sigma_{+}+\sigma_{-})
\end{equation}
in the Schroedinger picture, where the Pauli matrices associates with
the free energy and the transitions of the qubit and $E(t)=\mathcal{E}(t)\cos(\varphi-kx+\omega t)$
describes the electric field part of the dipole-field interaction
Hamiltonian. $\mu$ is the qubit dipole moment. For a resonant pulse,
the dressed states that diagonalize the Hamiltonian are
\begin{equation}
\left|\varepsilon_{\pm}(t)\right\rangle =\text{\ensuremath{\frac{1}{\sqrt{2}}\left[e^{-i\alpha(t)}\left|e\right\rangle \mp\left|g\right\rangle \right]}}\label{eq:dressed_st}
\end{equation}
where we have let $+$ ($-$) sign associate with the upper (lower)
energy state and $\alpha(t)=\varphi(t)-kx$ designate a time-dependent
phase factor. Correspondingly, we define $\nu_{-}=\left|\varepsilon_{-}\right\rangle \left\langle \varepsilon_{+}\right|$
and it Hermitian conjugate $\nu_{+}$ as an annhilation and creation
operator pair for the dressed states with eigenenergies $\Omega$
and $-\Omega$, respectively, where $\Omega(t)=\mu\mathcal{E}(t)$.

The bath is customarily written as a multimode resonator $H_{B}=\sum_{j}\omega_{j}\left(a_{j}^{\dagger}a_{j}\right)$,
from which the qubit-bath coupling is expressed by the Hamiltonian
\begin{equation}
H_{I}=\sum_{j}g_{j}\left(\left|e\right\rangle \left\langle g\right|+\left|g\right\rangle \left\langle e\right|\right)\left(a_{j}^{\dagger}+a_{j}\right).
\end{equation}
During the propagation of the microwave pulse, the qubit operators
in $H_{I}$ become the dressed operators associated with the basis
vectors of Eq.~\eqref{eq:dressed_st} and $H_{I}$ is correspondingly
transformed to $H_{I}(t)$ under the moving reference frame of the
pulse. In this frame, the total system-bath evolution under the interaction
picture is described by the Liouville equation
\begin{equation}
\frac{d\rho(t)}{dt}=-i\left[H_{I}(t),\rho(t)\right]\label{eq:Liouville_eq}
\end{equation}
for the density matrix $\rho=\rho_{\mathrm{S}}\otimes w$ of the density,
i.e. $\rho_{S}$ is the density matrix of the dressed qubit while
$w$ is the density matrix of the environment. Under typical considerations,
the environment is regarded as a reservoir with infinite energy supply
so it remains at a static density distribution $w$ during the course
of interaction while the environmental feedback affects the density
distribution of the system such that $\rho_{\mathrm{S}}=\rho_{\mathrm{S}}(t)$.
Henceforth, substituting the tensor product of density matrices into
Eq.~\eqref{eq:Liouville_eq} and tracing out the bath space, one
arrives at the master equation~\citep{ian18}
\begin{multline}
\frac{d\rho_{\mathrm{S}}(t)}{dt}=-i\left[H_{\mathrm{S}},\rho_{\mathrm{S}}\right]+\gamma\sin^{2}(\varphi-kx)\biggl\{\nu_{-}\rho_{\mathrm{S}}\nu_{+}\\
-\frac{1}{2}\left(\nu_{+}\nu_{-}\rho_{\mathrm{S}}+\rho_{\mathrm{S}}\nu_{+}\nu_{-}\right)\biggr\}\label{eq:master_eqn}
\end{multline}
where $\gamma(\Omega)=2\pi\sum_{j}g_{j}^{2}\delta(\omega_{j}-\Omega)$
represents the net decay rate stemmed from the bath spectral distribution.

The propagating microwave pulse is described by the Maxwell equation
\begin{equation}
\frac{\partial^{2}E}{\partial t^{2}}-c^{2}\frac{\partial^{2}E}{\partial x^{2}}=-\frac{1}{\epsilon_{0}}\frac{\partial^{2}P}{\partial t^{2}}\label{eq:Maxwell_eqn}
\end{equation}
where $P(t)=\left[\mathcal{P}(t)\exp i\{\varphi(t)+\omega t-kx\}+\mathrm{c.c.}\right]/2$
represents the macroscopic polarization of the qubit as a dipole.
We assume the same phase $\varphi(t)$ to ignore the calculation of
the dispersive effects and the complex amplitude correlates with the
density matrix through $\mathcal{P}=\mathrm{tr}\{\mu\sigma_{x}\rho_{\mathrm{S}}\}$.

\section{Propagation under dissipation\label{sec:Propagation}}

\subsection{The area equation}

In order to make the master-Maxwell equation pair \eqref{eq:master_eqn}-\eqref{eq:Maxwell_eqn}
solvable through decoupling, Eq.~\eqref{eq:Maxwell_eqn} is customarily
first reduced to first-order equations in the coordinate of local
time $\tau=t-x/v$
\begin{align}
\frac{d\mathcal{E}}{d\tau} & =\frac{\omega}{2(1-c/v)\epsilon_{0}}\Im\{\mathcal{P}\},\label{eq:enve_eqn}\\
\frac{d\varphi}{d\tau} & =-\frac{\omega}{2(1-c/v)\epsilon_{0}\mathcal{E}}\Re\{\mathcal{P}\},\label{eq:ph_eqn}
\end{align}
where the differential operator $\partial/\partial\tau$ contracts
from $\partial/\partial t+c\partial/\partial x$ up to a proportion
constant $(1-c/v)$. The reduction is made possible by assuming the
slow-varying envelope approximation, i.e. $\partial\mathcal{E}/\partial t\ll\omega\mathcal{E}$,
$\partial\mathcal{E}/\partial x\ll k\mathcal{E}$, etc., which omits
the second-order fast-varying terms. The local time $\tau$, which
is equivalent to the diagonal axis in the $xt$-plane, can also be
regarded as a time mark registered on the wavefront of the pulse. 

We have shown in Ref.~\citep{ian18} that the mixed qubit-field system
whose evolution is governed by Eq.~\eqref{eq:master_eqn} under the
dressed basis $\left|\varepsilon_{\pm}\right\rangle $ can be solved
by perturbative expansion, giving rise to $\mathcal{P}=\mu\left(1-e^{-\Gamma}-i\sin\theta e^{-\Gamma/2}\right)$
for an inital ground-state qubit, where 
\begin{equation}
\Gamma(\tau)=\int_{\tau_{0}}^{\tau}ds\,\gamma(\Omega)\sin^{2}(\varphi-kx)\label{eq:Gamma}
\end{equation}
denotes a decoherence factor and $\theta(\tau)=\int_{\tau_{0}}^{\tau}ds\,\Omega(s)$
denotes the enveloped area of the pulse up to time $\tau$. With the
determination of $\mathcal{P}$, the equation pair~\eqref{eq:enve_eqn}-\eqref{eq:ph_eqn}
becomes
\begin{align}
\frac{d\mathcal{E}}{d\tau} & =M^{2}\sin\theta e^{-\Gamma/2},\label{eq:enve_eqn_2}\\
\frac{d\varphi}{d\tau} & =\frac{M^{2}}{\mu\mathcal{E}}\left(1-e^{-\Gamma}\right),\label{eq:ph_eqn_2}
\end{align}
where $M=\sqrt{\mu^{2}\omega v/2(c-v)\epsilon_{0}}$. Note that in
this form, $\Gamma$, $\mathcal{E}$, and $\varphi$ are inter-dependent,
making the equation pair insolvable. Especially, the expression of
$\Gamma(\tau)$ demonstrates the memory effect of the culminated feedback
from the environment to the pulse. On one hand, it is determined by
the spectral distribution of the bath through $\gamma(\Omega)$; on
the other, it depends on the historic variation of the phase $\varphi$.

To give a realistic estimate of the environmental influence, the Leggett
model~\citep{caldeira81} is assumed, i.e. $\Gamma=\lambda\theta$
by regarding that the influence contributed by the phase through $\sin^{2}(\varphi-kx)$
averages out over the integration in Eq.~\eqref{eq:Gamma} and the
spectrum $\{g_{j}\}$ of the bath has a linear dependence on $\Omega$.
$\lambda$ here acts as a scale factor, for which Eq.~\eqref{eq:enve_eqn_2}
becomes the second-order equation
\begin{equation}
\ddot{\theta}=M^{2}e^{-\lambda\theta/2}\sin\theta.\label{eq:area_eqn}
\end{equation}
When $\lambda$ vanishes, the equation reduces to a typical pendulum
equation, for which a hyper-secant solution exists. This signifies
a solitary wave can travel absorption-free when encountering a qubit,
analogous to the effect of self-induced transparency (SIT) experienced
by a traveling light field through an ensemble of two-level atoms~\citep{mccall67}.

The extra factor $e^{-\lambda\theta/2}$ contributed by the environment
does not permit an explicit expression of the solution but does allow
an implicit solution. First, by transforming the variable from $\tau$
to $\theta$ in Eq.~\eqref{eq:area_eqn}, i.e. letting $\ddot{\theta}=\dot{\theta}(d\dot{\theta}/d\theta)=d(\dot{\theta}^{2})/2d\theta$,
the equation order is reduced by one. Then formally integrating by
parts over $\theta$ and taking the square root leads to the formula
\begin{equation}
\dot{\theta}=M\sqrt{\frac{2-e^{-\lambda\theta/2}(2\cos\theta+\lambda\sin\theta)}{1+\lambda^{2}/4}}.\label{eq:theta_deri}
\end{equation}
Since there is an one-one correspondence between $\theta$ and $\tau$
and the expression of $\dot{\theta}$ is differentiable with respect
to $\tau$, taking the reciprocal of $d\theta/d\tau$ and integrating
both sides with respect to $\theta$ again leads to the inverted function
\begin{multline}
\tau(\theta)=\tau_{0}+\frac{\sqrt{1+\lambda^{2}/4}}{M}\times\\
\int_{\theta_{0}}^{\theta}d\vartheta\left[2-e^{-\lambda\vartheta/2}(2\cos\vartheta+\lambda\sin\vartheta)\right]^{-1/2}\label{eq:tau_area}
\end{multline}
that shows the explicit dependence between $\theta$ and $\tau$,
where $\theta_{0}\to0$ at the limit $\tau_{0}\to-\infty$.

\subsection{Dissipation of pulse area}

Since the definition of $\tau$ is the time measurement from an apparatus
traveling at the wave speed $v$, one can regard the apparatus as
originally be placed at the pulse wavefront, where the pulse area
underneath the envelope is asymptotically zero. Then Eq.~\eqref{eq:tau_area}
as an implicit solution to the wave equation of Eq.~\eqref{eq:area_eqn}
should be intrepreted as expressing the time point read from the apparatus
when the underneath area culminates to a certain value $\theta$.

\begin{figure}
\includegraphics[clip,width=8.5cm]{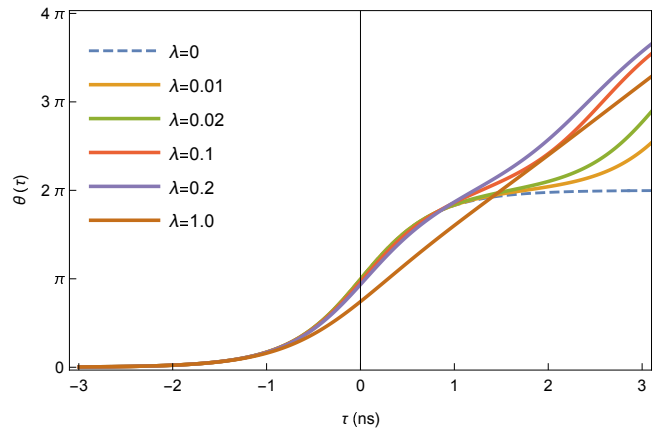}

\caption{Plot of the enveloped area $\theta(\tau)$ of a microwave pulse during
its propagation through a superconducting qubit against local time
$\tau$. The dashed curve indicates the scenario of zero decoherence
$\lambda=0$: $\theta(\tau)$ converges asymptotically to 0 and $2\pi$
towards the $-\infty$ and the $\infty$ time limits, respectively.
The variations of pulse area under the presence of decoherence are
given by the different color-coded solid curves (correspondence given
in the legend). For example, the red curve indicates the scenario
of finite decay with $\lambda=0.1$: $\theta(\tau)$ only converges
towards the $-\infty$ end and increases monotonically towards the
$\infty$ end.~\label{fig:area_evolution}}
\end{figure}

In Fig.~\ref{fig:area_evolution}, we plot the $\tau$-$\theta$
relation depicted in Eq.~\eqref{eq:tau_area} and graphically invert
the two reciprocal variables to make the relation appear intuitive.
The plots here and hereafter employ experimentally accessible parameters
extracted from current studies on superconducting circuit~\citep{eichler12,wen18,wen19}:
the qubit transition frequency $\omega_{z}/2\pi=5$GHz, the coupling
strength $\mu\mathcal{E}/2\pi=636$MHz at the pulse peak, and the
characteristic time $M^{-1}=0.5$ns that roughly estimates the pulse
width at half maximum. The latter is chosen according to typical solitonic
pulse generation, where the width is set to 10 cycles long of a resonant
microwave signal.

We observe that the zero-dissipation curve converges to a horizontal
asymptote at $2\pi$, showing that a steady-state solution exists
for a solitonic pulse of $2\pi$ enveloping area. The curve is anti-symmetric
about $\tau=0$ where $\theta(0)=\pi$. In other words, the time apparatus,
which originally travels in front of the incident pulse and has no
registered area at $\tau\to-\infty$, has begun to lag into the pulse
after it meets the qubit. It has lagged exactly halfway into the pulse
at $\tau=0$. For a symmetric-shaped solitonic pulse, this time point
coincides with the peak point of the pulse. These observations accord
with SIT~\citep{mccall67}, where microwave pulses with initial envelopping
area of $2n\pi$ can propagate through the qubit without being absorbed.

\begin{figure}
\includegraphics[clip,width=8.5cm]{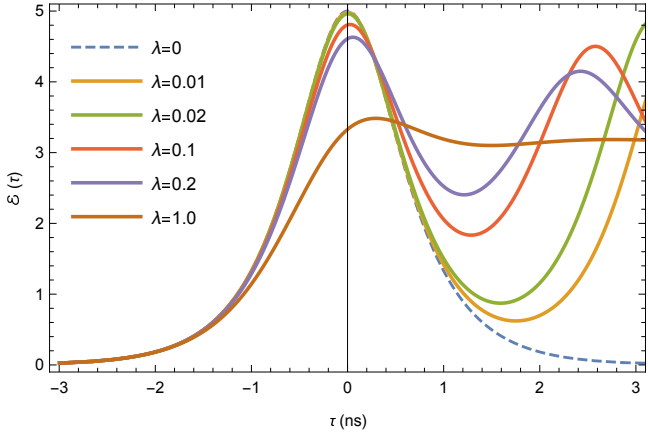}

\caption{The microwave pulse envelope $\mathcal{E}(\tau)$ as a function of
the local time $\tau$ during propagation under environmental dissipation
at different scale factor $\lambda$. The dashed curve corresponds
to $\lambda=0$ while the solid curves correspond to finite values
of $\lambda$, the color coding scheme being identical to that of
Fig.~\ref{fig:area_evolution}. The dissipative qubit splits up the
envelop in addition to absorbing the microwave photon, whereas the
non-dissipative qubit retains the solitonic shape of the pulse.~\label{fig:evlp_evolution}}
\end{figure}

Otherwise, with the presence of environmental dissipation, $\tau$
increases monotonically with $\theta$, showing that the longer the
pulse sustains its passage through the qubit, the more pulse area
is registered by the time apparatus, which is equivalent to energy
being extracted by either the qubit or the environment. When the scale
factor $\lambda$ remains small, the early development up to $\tau<1$ns
remains close to the case of $\lambda=0$. That is, when one regards
Eq.~\eqref{eq:area_eqn} as describing a Markovian process, the rate
of change of the system at an early stage is determined predominantly
by $\sin\theta$, not the environmental $\exp\{-\lambda\theta/2\}$,
when $\lambda$ is sufficiently small. During the early interaction,
the qubit is being inverted to its excited state and the dissipation
described by the Lindbladian in Eq.~\eqref{eq:master_eqn} would
make the first half of a $2\pi$-solitonic pulse insufficient to accomplish
total inversion at $\tau=0$. The observation is verified when we
plot in Fig.~\ref{fig:evlp_evolution} the envelope variation against
$\tau$ by taking the time derivative of $\theta$ numerically under
the same set of scale factors as given in Fig.~\ref{fig:area_evolution}.

Therefore, ever increasing the scale factor $\lambda$ leads the peak
point of a solution to Eq.~\eqref{eq:area_eqn} to deviate from $\tau=0$
and leans ever deeper into the right end. The dissipative environment
hence breaks the time symmetry of a permissible solution and forces
the qubit to fully invert only when $\tau>0$. When $\lambda$ is
sufficiently large (e.g. the brown curves in Figs.~\ref{fig:area_evolution}
and \ref{fig:evlp_evolution} with $\lambda=1$), the dissipation
becomes so fast that full inversion is never reached. Consequently,
apparent deviations from the $\theta=2\pi$ horizontal asymptote occur
during the second half of the qubit-pulse interaction. In the fully
(respectively, non-fully) inverted case, stimulated (respectively,
spontaneous) emission dominates while the qubit flips back to the
ground state. Further, if spontaneous emission dominates, $\theta$
follows rather a diagonal asymptote against $\tau$, which corresponds
to a leveled horizontal asymptote in $\mathcal{E}(\tau)$. This asymptotic
behavior means that when $\exp\{-\lambda\theta/2\}$ dominates $\sin\theta$
in Eq.~\eqref{eq:area_eqn}, the admissible solution to Eq.~\eqref{eq:enve_eqn_2}
for microwave propagation is a leveled pulse train (i.e. square pulse),
which constantly supplies energy to compensate the loss in dissipation
and sustain propagation.

In general, since the qubit excitation is already asymmetric about
$\tau=0$ (absorbs when $\tau<0$ and emits $\tau>0$) without the
dissipation, the dynamic symmetry breaking due to dissipation not
only attenuates the obtainable peak registered by the time apparatus,
but also unanimously produces $\theta>2\pi$ at the right $\tau\to\infty$
limit. This signifies that a pulse with an initial area less than
$2\pi$ cannot fully travel through the qubit and be registered by
the time apparatus. The longer the duration of time record, the larger
energy should the pulse carry before encountering the qubit.

For the particular cases where $\lambda$ remains small and $\sin\theta$
still dominates, $\theta(\tau)$ can approach a value either greater
or less than the diagonal $\theta$-$\tau$ asymptote during the later
part of propagation. It depends on the emission rate, which is measured
from the combined stimulated and spontaneous emissions into the circuit
waveguide, relative to $\lambda$. In addition, if we regard Eq.~\eqref{eq:area_eqn}
as expressing the derivative $\dot{\mathcal{E}}$, we find that the
extrema of $\mathcal{E}$ given in Fig.~\ref{fig:evlp_evolution}
should occur at $\theta=n\pi$. The straightforward cases are $n=0$
(the horizontal asymptote $\mathcal{E}=0$ at left end) and $n=1$
(the qubit is fully inverted near $\tau=0$). For $n=2$, the horizontal
asymptote is resumed only when $\lambda=0$ because the solitonic
pulse solution has its first absorbed half radiated in phase with
its second non-absorbed half, producing induced transparency and no
interference. Otherwise, the environment can interfere by absorbing
and re-emitting a stimulated emission photon after the first inversion,
thereby allowing the qubit undergo a second inversion process. Consequently,
asymptotic pulse area is not registered but further local extrema.

\section{Phase variation\label{sec:Phase}}

\begin{figure}
\includegraphics[clip,width=8.5cm]{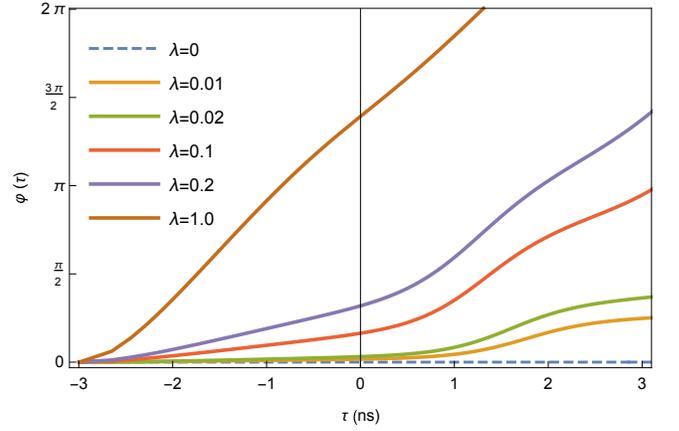}

\caption{The pulse phase $\varphi(\tau)$ as a function of the local time $\tau$.
The dashed line that corresponds to zero environmental influence serves
as a reference.~\label{fig:phase_evolution}}

\end{figure}

To obtain the solution of the phase variation, we first rewrite the
time derivate of $\varphi$: $\dot{\varphi}=\dot{\theta}(d\varphi/d\theta)$
where $\dot{\theta}=\mu\mathcal{E}$. Then the phase equation of Eq.~\eqref{eq:ph_eqn_2}
becomes
\begin{equation}
\frac{d\varphi}{d\theta}=\frac{M^{2}}{\mu^{2}\mathcal{E}^{2}}\left(1-e^{-\lambda\theta}\right).
\end{equation}
Substituting the expression of Eq.~\eqref{eq:theta_deri} and integrating
both sides over $\theta$, we obtain
\begin{multline}
\varphi(\theta)=\varphi_{0}+\left(1+\frac{\lambda^{2}}{4}\right)\times\\
\int_{\theta_{0}}^{\theta}d\vartheta\frac{\sinh(\lambda\vartheta/2)}{e^{\lambda\vartheta/2}-\cos\vartheta-(\lambda/2)\sin\vartheta}.\label{eq:phase_area}
\end{multline}
Juxtaposing Eq.~\eqref{eq:phase_area} against Eq.~\eqref{eq:tau_area}
for each data point value of $\theta$, we plot $\varphi$ as a function
of local time $\tau$ in Fig.~\ref{fig:phase_evolution} for a set
of different $\lambda$ values, where the color code again follows
that of Fig.~\ref{fig:area_evolution}.

We observe that the variation of the phase closely reflects the variation
of the enveloped area. At one extreme with $\lambda=0$, an SIT solitonic
pulse does not experience any phase change throughout the propagation,
as the vanishing $\lambda$ results in a zero RHS of Eq.~\eqref{eq:ph_eqn_2},
producing a constant $\varphi$. At the other extreme with $\lambda=1$,
the phase variation almost follows a diagonal asymptote in the $\varphi$-$\tau$
plane, similar to the area variation in the $\theta$-$\tau$ plane.
With a large $\lambda$, the exponential factor in Eq.~\eqref{eq:theta_deri}
approaches zero, leading to a constant $\mathcal{E}=\dot{\theta}/\mu=\sqrt{8}M/\lambda\mu$
as well as a constant slope $\dot{\varphi}=M\lambda/\sqrt{8}$ when
the propagated area $\theta$ is sufficiently large according to Eq.~\eqref{eq:ph_eqn_2}.
For other values of $\lambda$, we note two stages of phase variations
in general, the separating point of which follows the time point the
qubit obtains full inversion as discussed in the last section.

\section{Conclusions\label{sec:Conclusions}}

We have studied the variations of the envelope and the phase of a
microwave pulse during its propagation along a waveguide through a
qubit on a superconducting circuit. The study is given under the presence
of a dissipative environment which we have assumed to have a frictive
spectral distribution. Modelling on a coupled Maxwell-master equation,
we show that solution is admissible for one-shot pulses, where omission
of the environment reduces the solution to a symmetric soliton familiar
to classical SIT effects. The dissipative pulses have asymmetric and
multi-peak shapes, depending on the scale factor of the frictive environment.
We have given detailed analysis on these shapes in both the aspects
of envelope and phase. Such detailed knowledge would benefit the design
of more sophisticated pulses to control the qubit state for the purpose
of storing and processing quantum information.
\begin{acknowledgments}
Y.-B. Gao acknowledges the support of the National Natural Science
Foundation of China under Grant No.~11674017. H. I. acknowledges
the support by FDCT of Macau under grant 065/2016/A2, University of
Macau under grant MYRG2018-00088-IAPME, and National Natural Science
Foundation of China under grant No.~11404415.
\end{acknowledgments}


\begin{thebibliography}{99}
\bibitem{clarke08}J. Clarke and F. K. Wilhelm, Nature \textbf{453},
1031 (2008).

\bibitem{jqyou11}J. Q. You and F. Nori, Nature \textbf{474}, 589
(2011).

\bibitem{ian10}H. Ian, Y. Liu, and F. Nori, Phys. Rev. A \textbf{81},
063823 (2010).

\bibitem{martinis05}J. M. Martinis, K. B. Cooper, R. McDermott, M.
Steffen, M. Ansmann, K. D. Osborn, K. Cicak, S. Oh, D. P. Pappas,
R. W. Simmonds, and C. C. Yu, Phys. Rev. Lett. \textbf{95}, 210503
(2005).

\bibitem{martinis03}J. M. Martinis, S. Nam, J. Aumentado, K. M. Lang,
and C. Urbina, Phys. Rev. B \textbf{67}, 094510 (2003).

\bibitem{bialczak07}R. C. Bialczak, R. McDermott, M. Ansmann, M.
Hofheinz, N. Katz, E. Lucero, M. Neeley, A. D. O\textquoteright Connell,
H. Wang, A. N. Cleland, and J. M. Martinis, Phys. Rev. Lett. \textbf{99},
187006 (2007).

\bibitem{yoshihara06}F. Yoshihara, K. Harrabi, A. O. Niskanen, Y.
Nakamura, and J. S. Tsai, Phys. Rev. Lett. \textbf{97}, 167001 (2006).

\bibitem{ithier05}G. Ithier, E. Collin, P. Joyez, P. J. Meeson, D.
Vion, D. Esteve, F. Chiarello, A. Shnirman, Y. Makhlin, J. Schriefl,
and G. Schön, Phys. Rev. B \textbf{72}, 134519 (2005).

\bibitem{mallet09}F. Mallet, F. R. Ong, A. Palacios-Laloy, F. Nguyen,
P. Bertet, D. Vion, and D. Esteve, Nat. Phys. \textbf{5}, 791 (2009).

\bibitem{yu17}S. Yu, Y. Gao, and H. Ian, Quant. Inf. Process. \textbf{16},
283 (2017). 

\bibitem{tian02}L. Tian, S. Lloyd, and T. P. Orlando, Phys. Rev.
B \textbf{65}, 144516 (2002).

\bibitem{jqyou07}J. Q. You, X. Hu, S. Ashhab, and F. Nori, Phys.
Rev. B \textbf{75}, 140515 (2007).

\bibitem{mcdermott09}R. McDermott, IEEE Trans. Appl. Supercond. \textbf{19},
2 (2009).

\bibitem{caldeira83}A. O. Caldeira and A. J. Leggett, Ann. Phys.
\textbf{149}, 374 (1983).

\bibitem{leggett84}A. J. Leggett, Phys. Rev. B \textbf{30}, 1208
(1984).

\bibitem{wei06}L. F. Wei, Y. Liu, and F. Nori, Phys. Rev. Lett. \textbf{96},
246803 (2006).

\bibitem{neeley10}M. Neeley, R. C. Bialczak, M. Lenander, E. Lucero,
M. Mariantoni, A. D. O\textquoteright Connell, D. Sank, H. Wang, M.
Weides, J. Wenner, Y. Yin, T. Yamamoto, A. N. Cleland, and J. M. Martinis,
Nature \textbf{467}, 570 (2010).

\bibitem{caldeira81}A. O. Caldeira and A. J. Leggett, Phys. Rev.
Lett. \textbf{46}, 211 (1981).

\bibitem{widom82}A. Widom and T. D. Clark, Phys. Rev. Lett. \textbf{48},
63 (1982).

\bibitem{ian18}Y. Gao, S. Jin, and H. Ian, arXiv: 1811.05126 (2018).

\bibitem{basov66}N. G. Basov, R. V. Ambartsumyan, V. S. Zuev, P.
G. Kryukov, and V. S. Letokhov, Soviet JETP \textbf{23}, 16 (1966).

\bibitem{mccall67}S. L. McCall and E. L. Hahn, Phys. Rev. Lett. \textbf{18},
908 (1967).

\bibitem{eichler12}C. Eichler, C. Lang, J. M. Fink, J. Govenius,
S. Filipp, and A. Wallraff, Phys. Rev. Lett. \textbf{109}, 240501
(2012). 

\bibitem{wen18}P. Y. Wen, A. F. Kockum, H. Ian, J. C. Chen, F. Nori,
and I.-C. Hoi, Phys. Rev. Lett. \textbf{120}, 063603 (2018). 

\bibitem{wen19}P. Y. Wen, K.-T. Lin, A. F. Kockum, B. Suri, H. Ian,
J. C. Chen, S. Y. Mao, C. C. Chiu, P. Delsing, F. Nori, G.-D. Lin,
and I.-C. Hoi, Phys. Rev. Lett. \textbf{123}, 233602 (2019). 
\end{thebibliography}
\end{document}